\documentclass[twocolumn, times]{aastex631}

\newcommand{\aave}[1]{\left\langle #1\right\rangle}
\renewcommand{\vec}{\bf{}}
\usepackage{enumerate}

\begin{document}

\title{Energetically-dominant Sunward-Propagating Alfv\'en Waves Near 1 au and Their Relation to Large-scale Magnetic Switchbacks}

\author[0009-0001-4304-0809]{Nickolas Giardetti}
\affiliation{Department of Aerospace, Physics and Space Sciences, Florida Institute of Technology \\
150 West University Boulevard, Melbourne, FL 32901, USA}

\author[0000-0002-2358-6628]{Sofiane Bourouaine}
\affiliation{Department of Aerospace, Physics and Space Sciences, Florida Institute of Technology \\
150 West University Boulevard, Melbourne, FL 32901, USA}

\author[0000-0002-8841-6443]{Jean C. Perez}
\affiliation{Department of Aerospace, Physics and Space Sciences, Florida Institute of Technology \\
150 West University Boulevard, Melbourne, FL 32901, USA}



\begin{abstract}
In this letter, we investigate the population of energetically-dominant sunward-propagating Alfv\'en waves (SAWs) using more than 20 years of data provided by the~\emph{Wind} spacecraft near 1 au. We refer to SAWs as energetically-dominant sunward-propagating Alfv\'en waves within inertial range scales. Key parameters such as normalized cross helicity, plasma incompressibility, and magnetic incompressibility are used to determine the SAWs. Incorporating the polarity of the heliospheric magnetic field, AW modes are identified, which enables the determination of the propagation direction. Occurrence rates of SAWs vary from 1\% to 14\% depending on the time scale and solar wind stream type considered. Particularly, the relationship between large-scale magnetic field switchbacks (SBs) and SAWs (for a 1-hour long time scale) is investigated. A methodology utilizing pitch angle distributions of suprathermal electron strahl is employed to identify inverted magnetic field topology. The intervals containing SAWs are cross-referenced and examined with intervals identified as SBs. For a sample of 1636 1-hour SAW intervals, 17.5\% are associated with magnetic field switchbacks occurring at scales larger than one hour. The analysis lends support to the idea of switchbacks as one of the candidate sources for a portion of the SAW population.
\end{abstract}



\section{Introduction} \label{sec:intro}
The solar wind is a tenuous plasma that emanates from the Sun and permeates the heliosphere. It has been documented for decades that the solar wind is populated with Alfv\'enic fluctuations (e.g.~\cite{burlaga71}, \cite{belcher71}, \cite{volk75}). These Alfv\'en waves (AWs) are characterized by aligned or anti-aligned fluctuations in the plasma velocity and magnetic fields \citep{alfven42} with $\delta {\bf u}=\pm \delta {\bf b}$. Here, $\delta {\bf u} = \bf{u} - \bar{\bf{u}}$ represents the fluctuating plasma bulk velocity, and $\delta {\bf b} = \delta {\bf B}/\sqrt{\mu_0 \rho}$, represents the fluctuating Alfv\'en velocity. $\rho$ is the mass density and $\delta \bf{B} = \bf{B} - \bar{\bf{B}}$ is the fluctuating magnetic field. $\bar{\bf{u}}$ and $\bar{\bf{B}}={\bf B}_0$ are the mean values of the velocity and magnetic field, respectively. In situ, it has been observed that the most energetic Alfvénic fluctuations normally propagate away from the Sun (anti-sunward). Nonetheless, there is a rare subset in which AWs that propagate in the sunward direction are more energetic than anti-sunward fluctuations. Observation and documentation of sunward-propagating Alfv\'en waves (SAWs) that are \emph{energetically-dominant} has been sparse, and there has been a growing need for a more comprehensive survey.

Energetically-dominant SAWs were first observed in data from Voyager near 1.0 and 2.8 au, and noted to be a small portion of the overall cross helicity distribution~\citep{matthaeus1982}. More recently, individual SAW events have been observed on scales of 3, 12, and 16 hours with \emph{ACE}~\citep{gosling2009} and \emph{Wind}~\citep{li2016}, and have been shown to be associated with transient events such as magnetic reconnection and ion beams~\citep{he2015}. Despite their scarcity, identifying and characterizing SAWs is necessary, as they may be connected to important physical phenomena, such as magnetic reconnection~\citep{gosling2005, gosling2009, kigure2010}, magnetic field switchbacks (SBs)~\citep{bourouaine2020, schwadron2021, mallet2021} or due to AW turbulence cascades~\citep{perez09,perez12}. It is known that less energetic SAWs play an important role in models of solar wind turbulence~\citep{terres2024}. The interaction of counter-propagating Alfv\'en waves is known to be a required component for the initiation of turbulent cascades in the solar wind (\cite{bruno13}, \cite{tu95}), as well as the heating of fast solar wind plasma \citep{chandran09, cranmer2009,perez13} and slow solar wind \citep{damicis2019, bourouaine2024}. In this paper, we focus on the energetically dominant SAWs (hereafter referred to as SAWs). A comprehensive investigation of the occurrence of these waves and their sources may provide key insight into the mechanisms driving important heliophysical processes. As for the potential sources of SAWs, the primary phenomenon investigated herein is magnetic field inversions known as \emph{switchbacks}.

Switchbacks are transient events in the solar wind characterized by an abrupt reversal in magnetic field polarity~\citep{kasper2019, bale2019, dudok2020, bourouaine2020, jagarlamudi2023, raouafi2023}. The magnitude of the magnetic field is nearly constant, indicating the change in the field is purely rotational. Electron strahl also remains nearly constant, following the field as it bends \citep{halekas2020}. The S-shaped structures were first observed in the 1970s by Helios 1 and Helios 2~\citep{neubauer77}, then again by Ulysses in the 1990s \citep{balogh99}. More recent investigations with Parker Solar Probe suggest that switchbacks are likely generated from in-situ expansion \citep{tenerani21} and their characteristics are associated with the oscillations of large-amplitude AWs and the geometry of the background field \citep{bourouaine22}. Other sources of switchbacks may include magnetic flux ropes \citep{drake2021}, but specifically examining switchback formation is beyond the scope of this investigation. Work with PSP also has shown that AWs travel along the local field through switchback events \citep{mcmanus2020}. The net inward Els\"asser flux observed in \cite{mcmanus2020} shows that fluctuations during switchbacks are largely Alfv\'enic, thus it is highly likely there is a connection between switchbacks and SAW observations.

This paper presents the first systematic and most comprehensive analysis for identifying SAWs near 1.0 au using a large set of spacecraft data from \emph{Wind}. A statistical survey regarding the occurrence of SAWs near 1.0 au is provided. A methodology for identifying magnetic switchbacks is also presented. Confirmed events are cross-examined to identify if and when SAWs are present during switchbacks. The results of this investigation support the idea that switchbacks are a source for a considerable portion of the energetically-dominant sunward population of Alfv\'en waves.

\section{Identification of Sunward Alfv\'en Waves} \label{sec:alfven}
Data from \emph{Wind} is gathered from NASA's \emph{Space Physics Data Facility} (SPDF)~\citep{papitashvili20}. Estimates of density and velocity are obtained via by \emph{Wind's} SWE Faraday cup \citep{lin95}, and magnetic field data from its MFI fluxgate magnetometer \citep{lepping95}. Data from both instruments have measurement cadences of 24 seconds and span over 21 years of data from 2004 May 01 to 2025 June 12. All values are in the GSE coordinate system, wherein the positive $x$ direction is the Earth-Sun line.

The time series are discretized into non-overlapping intervals of window length $w \in \{1, 3, 6, 8, 12\}$ hours, and $w = 1$ hour is used as an example for the majority of the analysis. Each prospective interval is subjected to pre-processing filters, the first of which is a filter to identify and discard data gaps. It is ensured that each interval has less than 20\% missing data for each window size $w$. For intervals with a sufficient amount of data, the magnetic polarity is then examined. As we discuss below, one of the primary parameters for quantifying Alfv\'enicity is the normalized cross helicity, $\sigma_c$. The empirical estimation of cross helicity from spacecraft measurements (as well as residual energy and other plasma parameters) is sensitive to how the mean-field value of the magnetic field is calculated. As such, sufficiently mixed polarity of the instantaneous magnetic field may provide an inaccurate value for the local mean (average) magnetic field. To overcome that, we require at least 80\% of the data in each interval is a single polarity. When either criterion is not met, the interval is discarded entirely. After implementing the filters, we are left with a subset of intervals that have dominant magnetic polarities and no significant data gaps.

For the population of suitable intervals, three characteristic parameters are utilized to probe for Alfv\'enic fluctuations: the normalized cross helicity,
\begin{equation}
\sigma_c=2\frac{\aave{\delta\vec u\cdot\delta\vec b}}{\aave{\delta u^2 + \delta b^2}},
\label{eq:cross_helicity}
\end{equation}
the plasma compressibility,
\begin{equation}
c_n = \frac{\sqrt{\aave{(n-\bar{n})^2}}}{n_0},
\label{eq:plasma_compress}
\end{equation}
where $n$ and $\bar{n}$ are the instantaneous and mean plasma density, respectively, and the magnetic compressibility,
\begin{equation}
c_B = \frac{\sqrt{\aave{(B-\bar{B})^2}}}{\bar{B}},
\label{eq:mag_compress}
\end{equation}
\begin{figure*}[!t]
    \vspace{-.3cm}
    \includegraphics{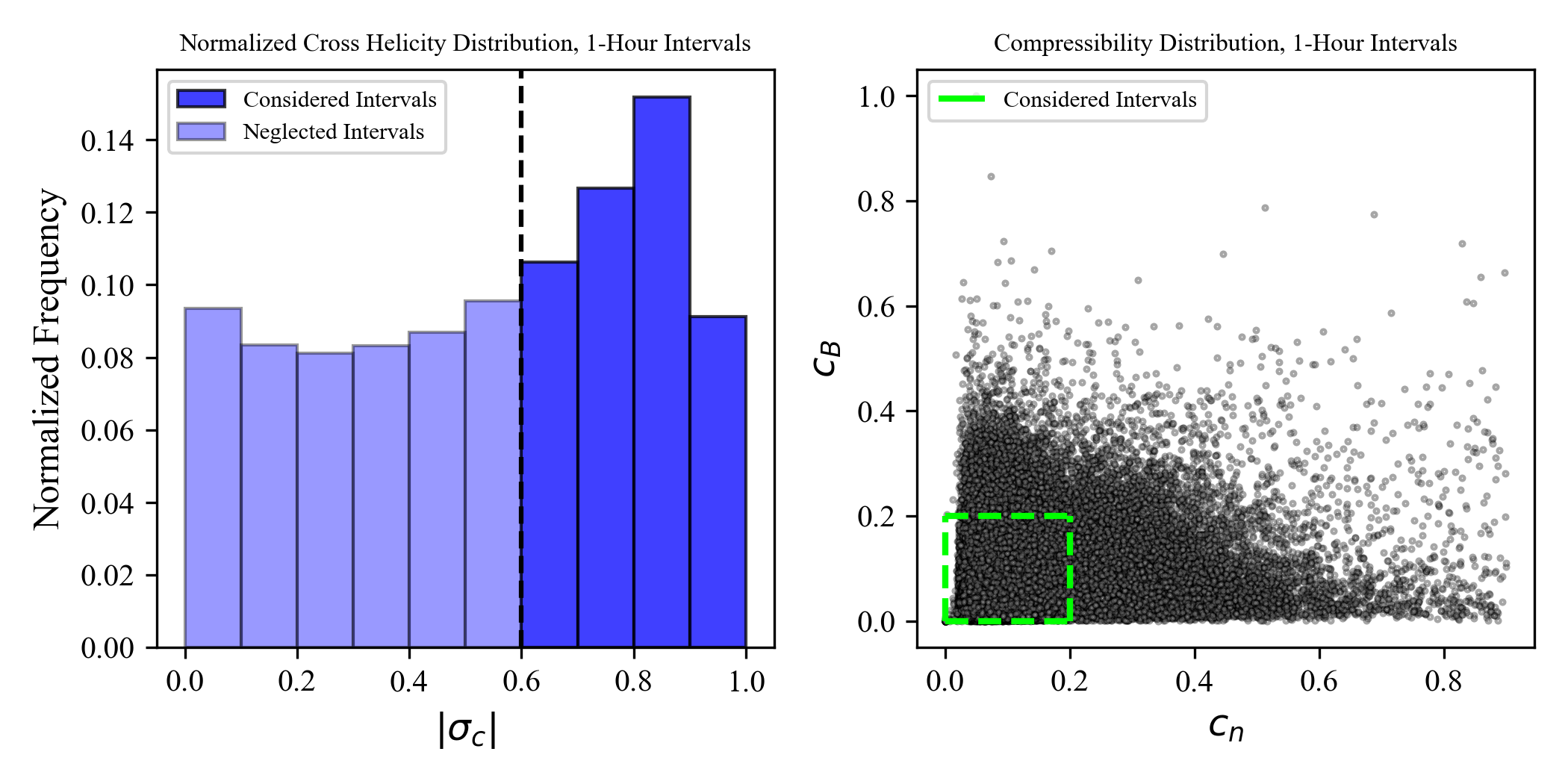}
    \caption{Selection criteria for Alfv\'enic intervals. Left Panel: Distribution of normalized cross helicity for all 1-hour intervals. The absolute value of $\sigma_c$ is considered as the sign merely indicates which Alfv\'en mode is dominant. The threshold of inclusion, $|\sigma_c| \geq 0.6$, is denoted by the dashed line. Right Panel: Distribution of plasma and magnetic compressibilities for all 1-hour intervals. The green square denotes included intervals satisfying $c_n, c_B \leq 0.2$. Intervals must satisfy \emph{both} $|\sigma_c| \geq 0.6$ and $c_n, c_B \leq 0.2$ to be considered for analysis.}
    \label{fig:thresholds}
\end{figure*}
where $B$ and $\bar{B}=\aave{B}$ are the magnitude of the instantaneous and the mean of the field strength, respectively. Here, the notation $\aave{\cdots}$ is used to denote a statistical ensemble average. Both $\bar{n}$ and $\bar{B}$ are calculated as the mean value of their respective parameters over the duration of the window. Equations (\ref{eq:cross_helicity}), (\ref{eq:plasma_compress}), and (\ref{eq:mag_compress}) are calculated for all intervals that met pre-processing requirements. 

The left panel of Figure~\ref{fig:thresholds} exhibits a wide range of $\sigma_c$ over all possible values $|\sigma_c|\le 1$ (for 1-hour intervals). When the cross helicity is close to zero, it indicates either a relative equipartition of energy between AWs, \emph{or} that Alfv\'enic fluctuations are not largely present. To ensure selected intervals have energetically-dominant Alfv\'en waves, we impose a threshold of $|\sigma_c| \geq 0.6$ (chosen as the value where the energy of one AW is four times larger than the other). Intervals with $|\sigma_c| \geq 0.6$ (with dark blue bars in Figure~\ref{fig:thresholds}) are then considered in our analysis.

Similarly, as solar wind data is examined through the framework of incompressible MHD, we require both compressibilities to meet: $c_n, c_B \leq 0.2$ to ensure incompressibility (shown in the right panel of Figure~\ref{fig:thresholds}). Other compressibility thresholds were checked ($c_B \leq 0.05, 0.1$), and did not significantly affect the statistics. Given the thresholds discussed, there are 187,992 non-overlapping intervals available for study. Of those, 144,188 met the pre-processing criteria (76.7\%). From that subset, 64,588 intervals remain after applying the Alfv\'enicity thresholds (34.4\% of the original data). The methodology ensures that only robust, refined intervals are used as the basis for analyzing the nature of SAWs. With the subset of Alfv\'enic intervals identified, the next step is to determine the direction of AW propagation.

The propagation direction for Alfv\'en waves is determined through the definition of the normalized cross helicity, and the polarity of the mean magnetic field, $\vec{B}_\mathrm{0}=\aave{\vec{B}}$. Thus, cross-referencing with the polarity of the magnetic field, four situations are possible:
\begin{enumerate}[(i)]
  \item $\sigma_c > 0$ , $B_{0x} > 0$ (anti-sunward AWs)
  \item $\sigma_c > 0$ , $B_{0x} < 0$ (sunward AWs)
  \item $\sigma_c < 0$ , $B_{0x} > 0$ (sunward AWs)
  \item $\sigma_c < 0$ , $B_{0x} < 0$ (anti-sunward AWs)
\end{enumerate}

\begin{figure*}[!t]
    \includegraphics[width=\linewidth]{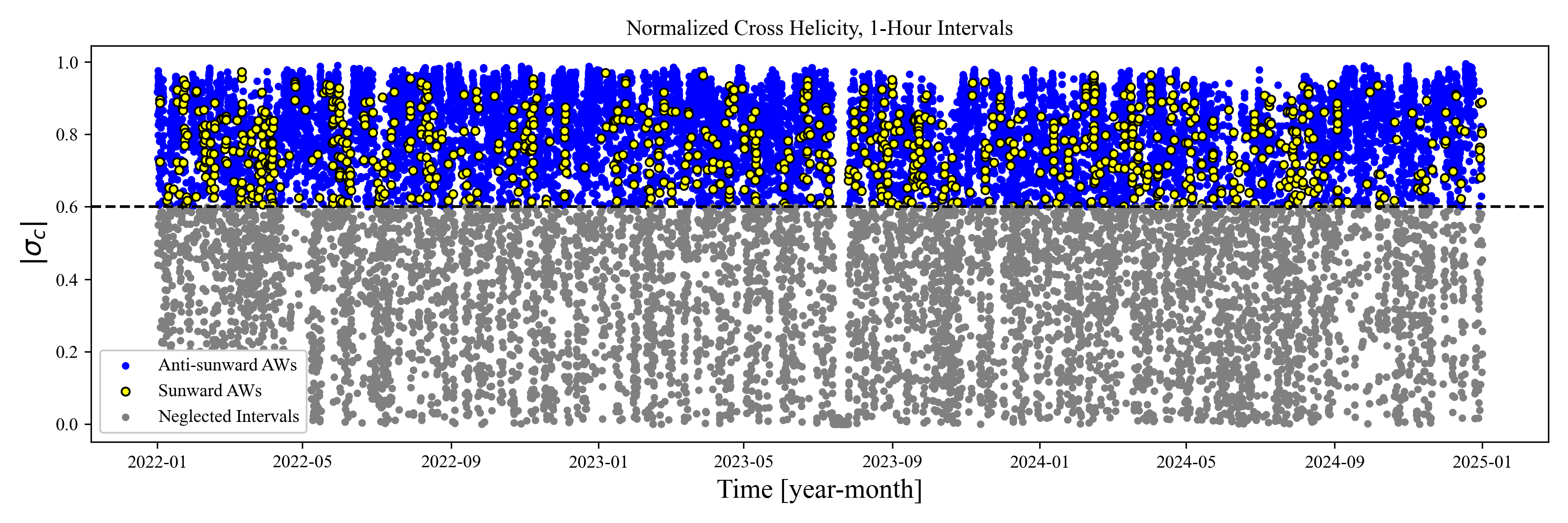}
    \caption{Sample of 1-hour intervals spanning 2022-2025, colored by propagation direction. Of the 18,264 intervals in the sample, 6,934 are filtered out (gray). The 998 sunward AWs, denoted in yellow, are relatively scarce compared to the 10,332 anti-sunward AWs (blue). All intervals shown meet the compressibility requirement $c_n, c_B \leq 0.2$.}
    \label{fig:sigma_scatter}
\end{figure*}
The sign of the normalized cross helicity provides information regarding the propagation direction with respect to the background field, and the polarity of the mean radial field, $B_{0x}$, then provides the propagation direction with respect to the Sun. The classification scheme is applied to each non-overlapping interval that meets the criteria discussed, creating populations of energetically-dominant Alfv\'enic fluctuations classified as sunward or anti-sunward. Figure~\ref{fig:sigma_scatter} shows the scatter plot of $\sigma_c$ values corresponding to intervals with SAWs in yellow circles and anti-sunward AWs (also referred to as ASAWs) in blue circles. The disregarded intervals are shown in gray circles.

Of the 64,588 energetically-dominant 1-hour AW intervals observed, 5705 intervals (8.8\%) are identified as SAWs. Table~\ref{tab:saw_stats} shows the occurrence of SAWs for energetically-dominant AW intervals of various window sizes in addition to 1-hour. Statistics for $w \in \{ 1, 3, 6, 8, 12\}$ hours are additionally partitioned by solar wind speed. Fast and slow streams with the same window size correspond to different characteristic length scales, as shown via Taylor's Hypothesis \citep{taylor1938} and the sweeping model of magnetohydrodynamics (MHD) \citep{bourouaine2019, bourouaine2020b}. Fast streams are taken as $u \geq 500$ $km s^{-1}$, and slow streams as $u<500$ $km s^{-1}$.
\begin{table}
    \centering
    Energetically-dominant AW Occurrence
    \begin{tabular}{ccccccc}
        \hline
        \hline
         &\multicolumn{2}{c}{All Streams}&  \multicolumn{2}{c}{Fast Streams}&  \multicolumn{2}{c}{Slow Streams}\\
         \hline
        $w$ [hr]&SAW&ASAW&  SAW&ASAW&  SAW&ASAW\\
        \hline
        1&5705&58883
&1086&20292&4619&38591\\
        3&815&16757
&128&6280&687&10477\\
        6&168&6979
&23&2750&145&4229\\
        8&76&4604
&4&1828&72&2776\\
        12&26&2496&2&1015&24&1481\\
    \end{tabular}
    \caption{Statistics of energetically-dominant AW modes for various window sizes split by solar wind stream type. While the minority across all scales, SAWs are more common in Alfv\'enic slow wind compared to fast wind. Additionally, energetically-dominant SAWs are more common at smaller scales.}
    \label{tab:saw_stats}
\end{table}

\begin{figure}
    \includegraphics[width=\linewidth]{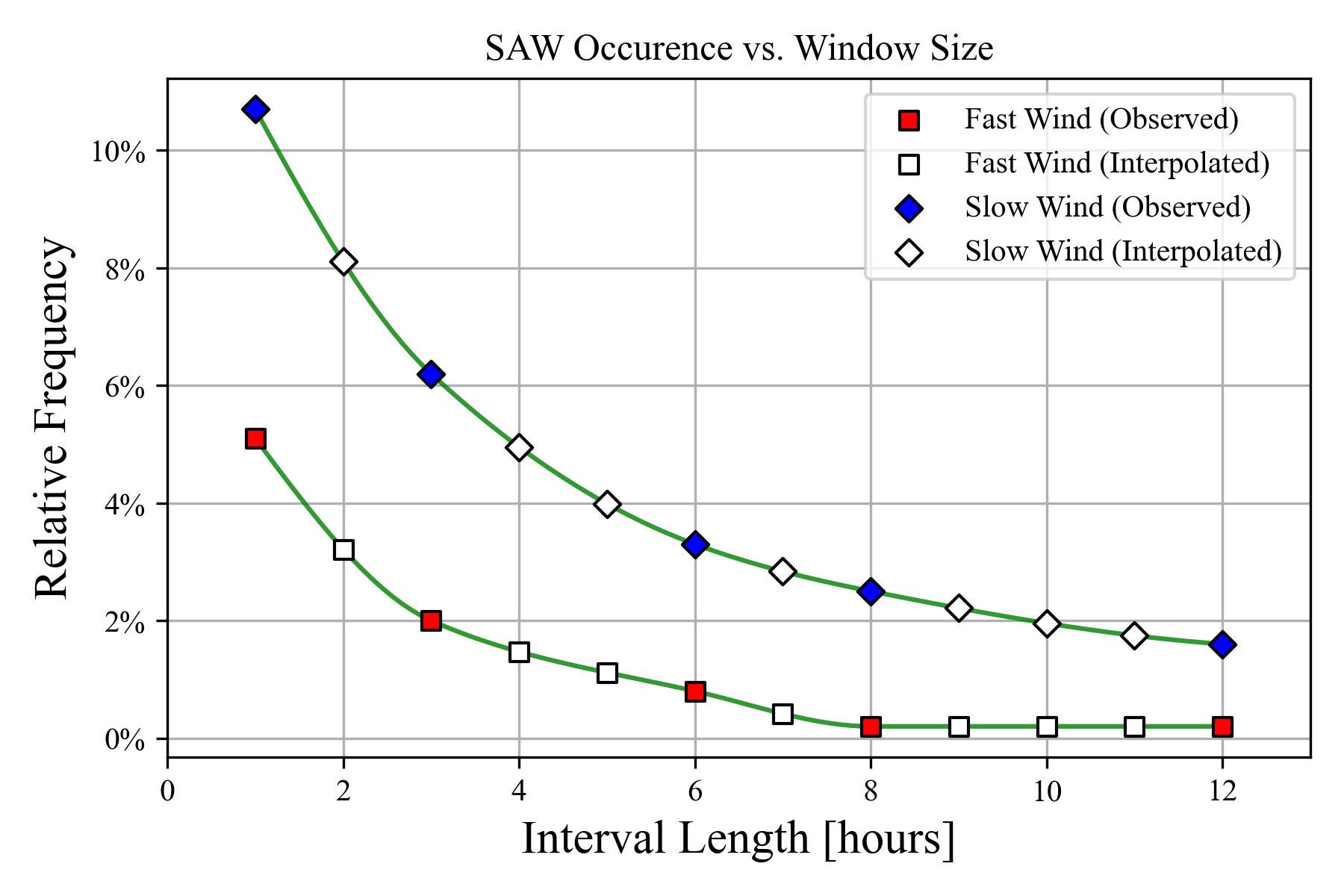}
    \caption{Occurrence of SAW as a function of window size. Slow solar wind streams (diamonds) contain more SAWs when compared with fast wind (squares). White markers are interpolated window sizes using cubic Hermite spline fitting.}
    \label{fig:SAW_occurrence}
\end{figure}

Figure~\ref{fig:SAW_occurrence} shows the normalized results of Table~\ref{tab:saw_stats}, with intermediate scales interpolated. For both stream types occurrence is observed to be decreasing with increasing window size $w$, although slow wind exhibits marginally more SAWs than fast wind. The result is plausible, as smaller windows equate to smaller spatial scales via Taylor's Hypothesis. Magnetohydrodynamic (MHD) numerical simulations have shown that the turbulent cascade in fully developed MHD turbulence can produce localized spatial patches—within the inertial range—where either parallel or anti-parallel Alfv\'enic (AW) turbulent fluctuations dominate \citep{perez09}. Consequently, solar wind turbulence may contain localized regions of SAWs embedded within a globally anti-sunward Alfv\'enic (ASAW) turbulent background. Furthermore, large-amplitude Alfv\'en waves are generally believed to originate from the Sun, whereas smaller-scale Alfv\'enic fluctuations are generated locally within the solar wind. The subset of 1-hour intervals identified as SAWs is the data used in subsequent analysis regarding large-scale magnetic switchbacks.

\section{Identification of Magnetic Switchbacks} \label{sec:switchbacks}
In our analysis, we are interested in large-scale magnetic switchbacks that last more than one hour. Magnetic field inversions exhibit a variety of data signatures, and some of them display high Alfvénicity \citep{horbury2018,kasper2019,bale2019,bourouaine2020,bourouaine22}. We primarily use electron pitch angle distributions (PADs), which are also obtained from \emph{Wind's} SWE instrument with a cadence of $\sim$12 seconds \citep{ogilvie21}.

To locate switchbacks associated with SAW intervals, the methodology of \cite{owens13} is adopted. The method relies on the concept that suprathermal electron strahl always follows magnetic field lines, regardless of polarity. As observed by the Parker Solar Probe, if strahl electrons propagate away from the Sun, the only instance where a sustained sunward electron strahl flow would exist is when the magnetic field has folded \citep{halekas2020}. As such, we can utilize the pitch angle distributions of strahl electrons to identify field inversions. Among the various PAD energy channels available, the 290.1 eV channel is chosen because it is outside the range of the core electron population while still having sufficient counts. Pitch angles are calculated with respect to nine-second averaged measurements of the instantaneous magnetic field. To determine where the dominant strahl electron flux exists, a background flux is established via the mean of the four pitch angle bins nearest to 90$^{\circ}$ (81$^{\circ}$, 87$^{\circ}$, 93$^{\circ}$, 99$^{\circ}$). The parallel and anti-parallel fluxes are determined from the four pitch angle bins nearest to 0$^{\circ}$ (3$^{\circ}$, 9$^{\circ}$, 15$^{\circ}$, 21$^{\circ}$) and 180$^{\circ}$ (177$^{\circ}$, 171$^{\circ}$, 165$^{\circ}$, 159$^{\circ}$), respectively. Following \cite{owens13} methodology, the background flux must be exceeded by at least 100\% for a strahl in either direction to be deemed present. If and when a definitive strahl is found, it is compared with the local magnetic field direction. In situations where the PAD has dominant flux either: A) parallel to a sunward field, or B) anti-parallel to an anti-sunward field, it can be deduced that a sunward-propagating strahl is present. The converse of the situations presented are classified as anti-sunward strahl, and situations where both sunward and anti-sunward strahl exist are classified as counterstreaming.

\begin{figure}
    \includegraphics[width=\linewidth]{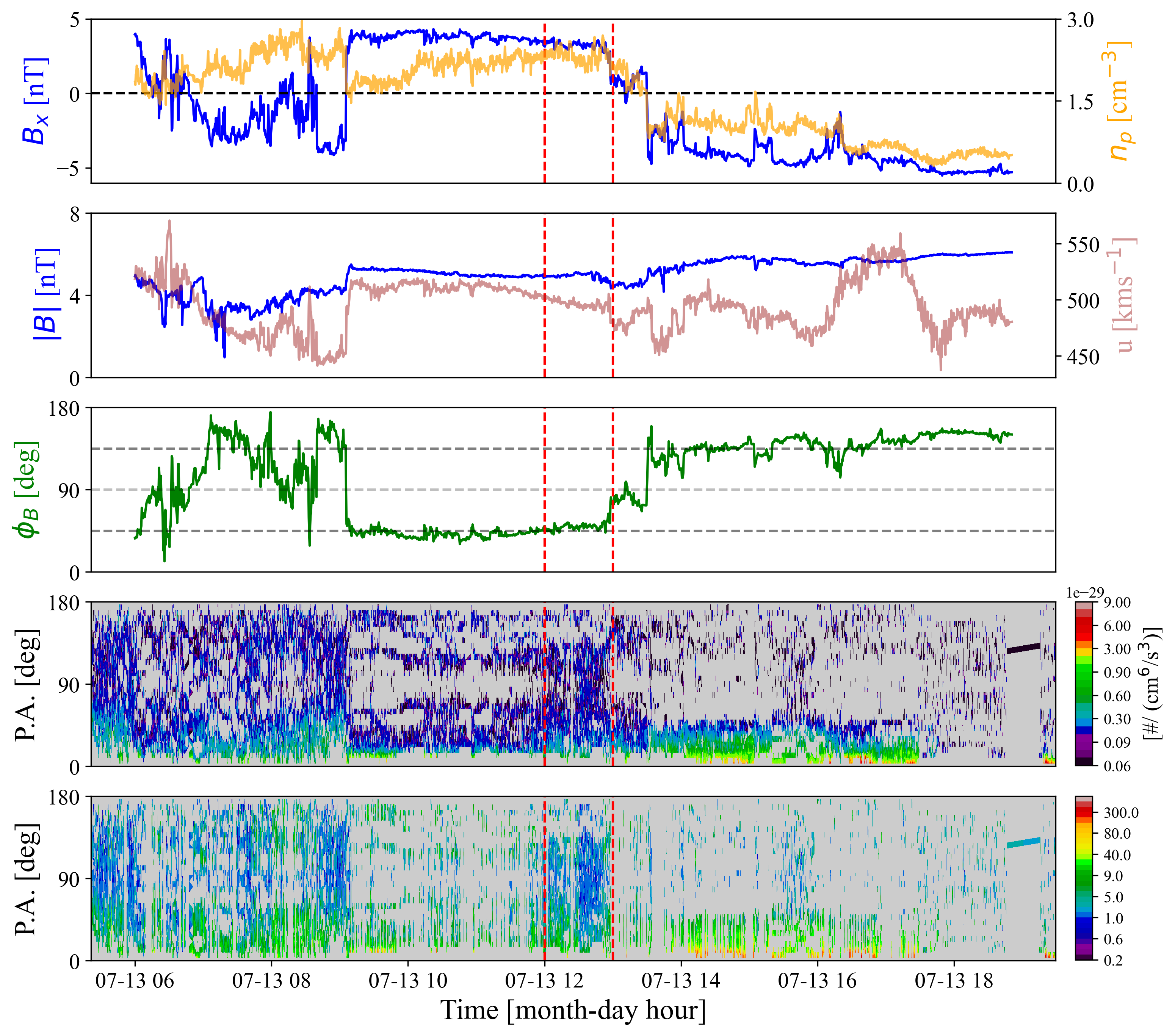}
    \caption{SAW parameters during an observed switchback in the inverted magnetic topology. The original 1-hour interval is centered within the red dashed lines, with six-hour neighbor windows on either side. Panel 1) radial component of the magnetic field (blue) and proton density (orange); Panel 2) magnitude of the magnetic field (blue) and plasma bulk velocity (light brown); Panel 3) magnetic radial angle; Panel 4) electron strahl PAD; Panel 5) electron strahl PAD normalized to the background flux. The gray lines in Panel 3 represent the Parker Angle near 1 au (45$^{\circ}$ and 135$^{\circ}$), along with 90$^{\circ}$.}
    \label{fig:combined_1}
\end{figure}

Counterstreaming strahl may occur due to closed magnetic field lines, or other transient events such as magnetic reconnection exhausts and ICMEs \citep{shodhan2000, anderson2012,borovsky2021}. Intervals where neither direction of strahl is present or the background flux is not exceeded are classified as undetermined. Undetermined intervals require additional methods to probe for switchbacks, and are to be addressed in a separate publication. While PAD data is available on the order of 12 seconds, strahl classifications are resampled to a time resolution of 24 seconds to match the plasma and field sampling rates. Utilizing the strahl categorization and the polarity of the magnetic field, the local magnetic topology is determined. The topology classifications are: 1) uninverted field, 2) inverted field, and 3) counterstreaming strahl. The 24-second topologies are aggregated to 1-hour cadence, where the most prevalent classification (the mode) is taken to be the dominant behavior. From the resulting classification scheme, any SAWs in inverted field intervals are considered to be part of large-scale switchbacks. Figure~\ref{fig:combined_1} shows an example of a switchback in the inverted category, and Figure~\ref{fig:saw_pies} shows the results of the topology classification.

\section{Discussion} \label{sec:discussion}
It is believed that the SAWs in inverted field intervals are catching the folded portions of switchbacks as found in \cite{bourouaine2020}. As such, by viewing the intervals adjacent to the inverted field intervals, the entire switchback can be observed. Figure~\ref{fig:combined_1} shows a SAW interval associated with a large-scale switchback from the inverted topology, with a normalized cross helicity of -0.66. The original interval considered is bounded by the dashed red lines. Upon expanding the interval on either side by six hours, it becomes apparent that the SAW interval is ``on top of" a field inversion. The radial component of the magnetic field switches from negative to positive (sunward in GSE coordinates), while the magnitude of the magnetic field remains constant. The plasma density remains fairly homogeneous, an enhancement in the bulk velocity is observed, and the magnetic radial angle (defined as $\phi_B = arccos(B_x/\bar{B})$) flips from anti-Parker-aligned to Parker-aligned. Panels 4 and 5 show that the PAD (raw and normalized, respectively) exhibits parallel strahl throughout the inversion.
\begin{figure}
    \includegraphics[width=\linewidth]{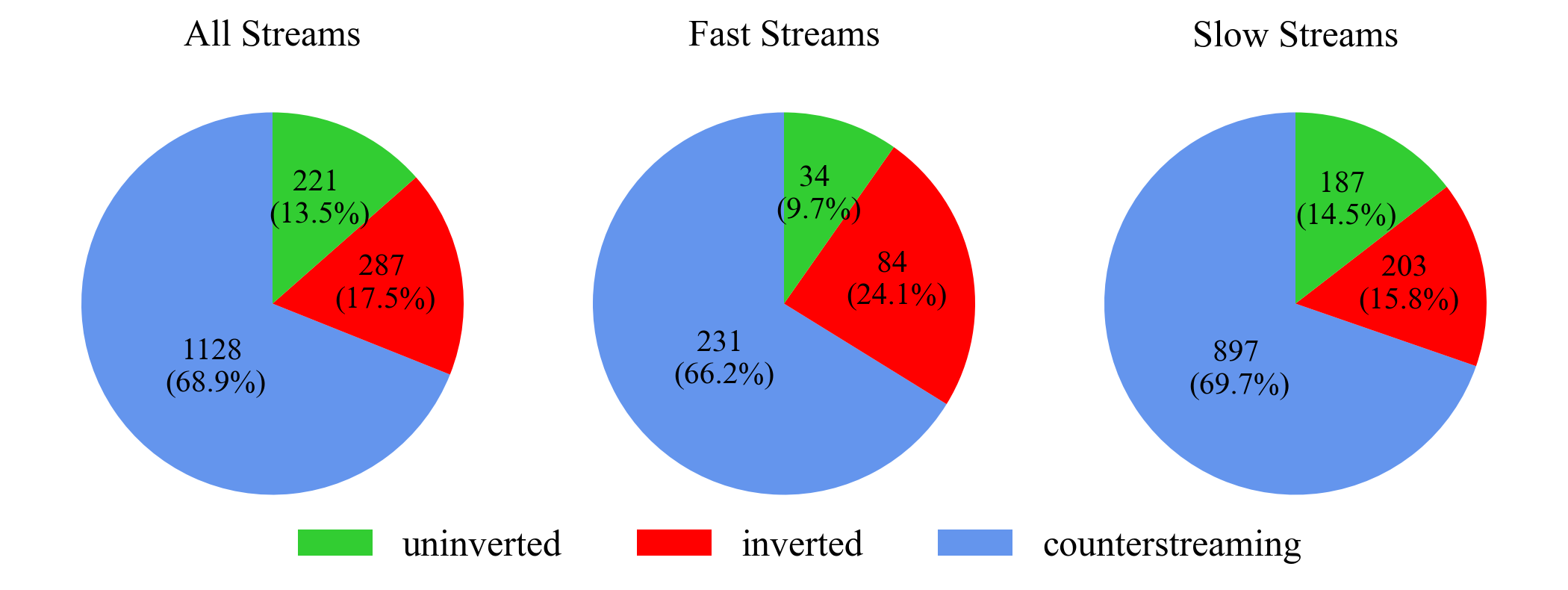}
    \caption{Distributions of magnetic topology split by solar wind stream for 1-hour intervals of energetically-dominant SAW. Counterstreaming strahl is indicated in blue. Uninverted and inverted topologies are represented by green and red sectors, respectively.}
    \label{fig:saw_pies}
\end{figure}

The statistical results of the magnetic topology within 1-hour SAW intervals are shown in Figure~\ref{fig:saw_pies}. The classification scheme is presented both in totality and split by solar wind stream type. Of the 5705 energetically-dominant SAW intervals, 1636 have sufficient PAD data to make topology classifications. While this represents approximately 29\% of the total SAW intervals, the statistics are sufficient to consider them a representative sample. Counterstreaming intervals are the most common, with an occurrence rate of roughly 69\%. Slow streams exhibit marginally more (3.5\%) counterstreaming intervals compared to fast streams. The counterstreaming intervals identified are recorded for examination in subsequent studies, as they are believed to be signatures of magnetic reconnection events that harbor SAWs. Inverted fields (switchbacks) are the next most common classification, comprising 17.5\% of the sample. Additionally, the inverted topology is 8.3\% more common in fast streams compared to slow streams. Uninverted intervals are the least common across both stream types, which supports the notion of energetically-dominant SAWs being rare in pristine solar wind. SAWs that are not associated with switchbacks (and do not exhibit counterstreaming strahl) require investigation for other potential sources such as stream interaction regions (SIRs), rotational discontinues (RDs), and turbulence. The interval and distribution of magnetic topologies shown in Figure~\ref{fig:combined_1} and Figure~\ref{fig:saw_pies} suggest a significant portion of energetically-dominant SAWs are associated with large-scale switchback events.

\section{Conclusion} \label{sec:conclusion}
This analysis presents a methodology for thoroughly identifying energetically-dominant sunward-propagating Alfv\'enic fluctuations. The methodology is implemented on $\sim$21 years of plasma and magnetic field data from the $Wind$ spacecraft to obtain the first systematic and most comprehensive characterization of SAW in the solar wind to date. The occurrence rates of SAWs show dependence on both fast and slow solar wind stream type and characteristic scale, with large-scale and fast wind SAWs being the most uncommon. Our analysis shows that slow wind exhibits slightly more SAWs than fast wind, but occurrence in both stream types decreases as window size is increased.

As large-scale magnetic switchbacks are considered to be a candidate source for SAWs, a methodology from \cite{owens13} is employed to identify magnetic topologies. For a large sample of 1-hour SAW intervals, counterstreaming strahl is the most common classification, followed by inverted field, and lastly uninverted field. The inverted topology, taken to be SAWs associated with large-scale switchbacks, occurs more frequently in fast wind intervals compared to slow wind. Overall, SAWs have been currently identified in three distinct situations: 1) associated with magnetic switchbacks; 2) associated with counterstreaming strahl; 3) not associated with either.

In future work, the investigation conducted at 1.0 au will be extended to the inner heliosphere, namely with Parker Solar Probe and Solar Orbiter. Some radial evolution of SAW occurrence was examined by \cite{li2016} for distances from 1-6 au (via Voyager). Now, with the plethora of inner-heliosphere data available via PSP and Solar Orbiter, the radial evolution of SAWs can be examined with more precision than past missions. Switchback events have also been readily observed in the inner heliosphere, and will be invaluable tools to continue understanding the mechanisms behind SAW generation. The prevalence of counterstreaming strahl occurring alongside SAWs also necessitates an investigation into a possible relationship with magnetic reconnection exhausts. The remaining intervals will be examined for evidence of other transient events as potential sources. This study and those in the future continue to enhance our understanding of SAWs, and the role they play in solar wind turbulence and heating.

The authors acknowledge support from NASA grants 80NSSC23K0776, 80NSSC24K0137, 80NSSC24K0564, and 80NSSC23K0419.

\bibliographystyle{aasjournal}

\end{document}